# Social Credibility incorporating Semantic Analysis and Machine Learning: A Survey of the State-of-the-Art and Future Research Directions

Bilal Abu-Salih[1,2], Bushra Bremie[1], Pornpit Wongthongtham[1], Kevin Duan[1], Tomayess Issa[1], Kit Yan Chan[1], Mohammad Alhabashneh[1], Teshreen Albtoush[1], Sulaiman Alqahtani[1], Abdullah Alqahtani[1], Muteeb Alahmari[1], Naser Alshareef[1], Abdulaziz Albahlal[1]

[1] Curtin University, Australia
{bilal.abusalih,bushra.bremie,ponnie.clark, Abdulaziz.albahlal, tomayess.issa, kit.chan, mohammad.alhabashneh, yuchao.duan}@curtin.edu.au, {bushra.bremie, Teshreen30}@gmail.com

[2] The University of Jordan, Jordan

**Abstract.** The wealth of Social Big Data(SBD) represents a unique opportunity for organisations to obtain the excessive use of such data abundance to increase their revenues. Hence, there is an imperative need to capture, load, store, process, analyse, transform, interpret, and visualise such manifold social datasets to develop meaningful insights that are specific to an application's domain. This paper lays the theoretical background by introducing the state-of-the-art literature review of the research topic. This is associated with a critical evaluation of the current approaches, and fortified with certain recommendations indicated to bridge the research gap.

## 1 Introduction

The dramatic increase in the social data impact - a testimony to our growing digital lifestyles - has taken on industries and activities ranging from marketing and advertising to intelligence gathering and political influence. Interpreting such a Big Data island is momentous to bring new perspectives and to improve business practices, yet this revolution is still in its infancy. It is easy to assume that social data revolution is about quantity, but social data is more than just volume. In fact, the extents of this revolution are more wide-spreading: it is about building data infrastructures that are needed to effectively digest the breeding of social data to infer the hoped-for added value. This has motivated the research communities to dig deep, provide solutions and to implement platforms into potential usage of these datasets to benefit several applications[1-7].

This paper presents a thorough survey of the state-of-the-art approaches drawn from academic sources with the relevance to the topic of this paper. The purpose of the review is threefold: Social Trust, Semantic Analysis, and Data Classification in the era of SBD.

## 2 SBD Incorporating Trust

In modern enterprises, social networks are used as part of the infrastructure for a number of emerging applications such as recommendation and reputation systems. In such applications, trust is one of the most important factors in decision making. Sherchan et al.[8] defined Trust as the measurement of confidence where a group of individuals or communities behave predictably.

### 2.1 Generic-based Trustworthiness Approaches

Generic-based trust approaches in Online Social Networks are those frameworks, techniques, and tools developed to calculate and infer the trustworthiness values of users and/or their content with no consideration to the domain(s) of interest which can be extracted from the user level or post level. The trustworthiness of social media data is now a crucial consideration. With such a vast volume of data being interchanged within the social media ecosystems, data credibility is a vital issue, especially regarding personal data [9].

**Graph-based Social Trust:** Podobnik et al. [10] propose a model that calculates trust between friends in a network graph based on weights of the edges between user's connected friends in Facebook. Agarwal and Bin [11] suggest a methodology to measure the trustworthiness of a social media user by using a heterogeneous graph in which each actor in the Twitter domain was presented as a vertex type in the graph. The level of trustworthiness was measured using a regressive spread process. On the other hand, the paper, neglects to consider the importance of weighting scheme and time factor. Each edge category should be assessed at different credibility levels; hence, a weighting scheme should be used. Trustworthiness values differ over time; consequently, the temporal/time factor should be integrated.

**Trust for recommendation systems:** Massa et al. [12], show a web of trust as an alternative to the standard way of ranking a user, i.e. standard recommendation systems. Further, Gupta et al. [13] present the "WTF: Who To Follow" service which is being used as a recommendation system for the Twitter social network. This service is used mainly as a recommendation driver and has a significant impact; by using it, numerous new connections have been created. Further work by Gallege et al. [14] and Sun et al. [15] propose trusted-based recommendation techniques.

**Trust incorporating Sentiment Analysis**: The use of sentiment analysis techniques to analyse the content of OSNs has significantly influenced several aspects of research. In the context of social trust, Alahmadi et al. [16] propose a recommendation system framework incorporating implicit trust between users and their emotions. AlRubaian et al. [17] present a multi-stage credibility framework for the assessment of microbloggers' content. The development of sentiment-based trustworthiness approaches for OSNs is discussed further in [18-20].

### 2.2 Domain-based/ Topic-Specific Trustworthiness Approaches

Adding a user-domain dimension when calculating trust in social media is an important factor, to enhance understanding of users' interest. In this context, the notion of domain-based trust for the data extracted from the unstructured content (such as social media data) is significant. This is through calculating trustworthiness values which correspond to a particular user in a particular domain. The literature of trust in social media shows a lack of approaches for measuring domain-based trust. Several reviews have carried out to highlight the importance of conducting a fine-grained trustworthiness analysis in the context of SBD [8, 21-23]. In particular, measuring the user's trustworthiness in each domain of knowledge is vital to better understanding users' behaviours in the OSNs.

## 3 SBD Incorporating Semantic Analysis, and Machine Learning for Data Classification and Topic Distillation

In the latter part of the 20th century, researchers in the field of Artificial Intelligence (AI) have become active in computational modelling of statistical analysis means alongside with defining ontologies that would deliver automated reasoning capabilities. This section aims to review the existing approaches into two main categories: (i) techniques incorporating semantic analysis for domain discovery; (ii) machine learning statistical techniques for data classification topics distillation.

### 3.1 SBD Incorporating Semantic Analysis

The Semantic Web (SW) was introduced by Berners Lee who provided a new vision for the next web where data is given semantic meanings via data annotation and manipulation in a machine-readable format [24]. Ontology-based on Gruber's [25] definition is the formal explicit specification of a shared conceptualization within a domain, as a form of concepts and relationships between these concepts which is used to describe the domain. By incorporating semantic analysis, semantic data can be inferred from social media data. The use of ontology in the social media has been applied widely to infer semantic data in a broad range of applications. De Nart et al. [26] propose a content-based approach to extract the main topics from the tweets. This approach is an attempt to understand the research communities' activities and their emerging trends. Chianese et al. [27] propose a data-driven and ontology-based approach to identify cultural heritage key performance indicators as expressed by social network users. This approach can be used in different domains but is only relevant or ad-hoc to user domains. Michelson and Macskassy [28] use the DBpedia knowledge base to annotate entities in users' tweets, and extract the users' main interests by using the categories proposed in Wikipedia. Wikipedia as a knowledge base repository has been utilised for topic discovery in [29, 30].

## 3.2 SBD Incorporating Machine Learning for Data Classification and Topic Distillation

Machine Learning applications enable real-time predictions leveraging high quality and well-proven learning algorithms. Based on the current dominant position and high impact on business in several use cases, according to Gartner's recent report on emerging technologies[1], incorporating machine learning, in particular, enhances the decision-making process and provides valuable insights from large-scale data. Topic distillation (a.k.a topic discovery, or topic modelling, or latent topic modelling or statistical topic modelling) is an automatic approach to distil topics from a corpus of words embodied in a set of documents incorporating statistical techniques [31-33].

These statistical-based techniques have been used as another means of topic modelling and discovery in social data mining. Examples of such statistical-based techniques that have been used are LDA (Latent Dirichlet Allocation) [34], Latent Semantic Analysis (LSA), and recently Fuzzy Latent Semantic Analysis (FLSA)[35]. LDA is based on an unsupervised learning model to identify topics from the distribution of words. In LSA, an early topic modelling method has been extended to pLSA [36], which generates the semantic relationships based on a word-document co-occurrence matrix. FLSA supposes that the list of documents and their embodied words can be fuzzy clustered where each cluster can be represented by a certain topic. LDA and similar unsupervised techniques have been widely used in several modelling applications [37-42].

## 4 Critical Synthesis Review of the Current Approaches

The previous sections present an integrated review of the existing methods and approaches in the era of SBD incorporating trust, semantic analysis and machine learning.
- Lack of advanced domain-based trustworthiness approaches,
- Lack of managing and extracting high-level domains from the textual content of SBD,
- Lack of domain-based techniques for dual classification.

### 4.1 Lack of an Advanced Domain-based Trustworthiness Approaches

*A need for domain-based trustworthiness*: There have been some efforts apply generic-based credibility evaluation approaches for users and their content in OSNs [43-45]. However, they do not take the topic or subject factor into consideration; the classification has been computed in general. Users will have a certain reputation in one domain, but that does not always apply to any other domain. The user's credibility should be domain-driven. For example, evaluating users' trustworthiness in a specific domain has been driven by its implication in several applications such as personalized recommendation systems [44], opinion analysis [46], expertise retrieval [47], and computational advertising [48].

---

[1] http://www.gartner.com/document/3383817?ref=solrAll&refval=175496307&qid=34ddf5254 22cc71383ee22c858f2238a, Visited in 25/10/2016.

***Lack of Incorporating Temporal Factor***: Subsequent studies have focused on the users' topic(s) of interest or their domains of knowledge. However, studying users' behaviour over time has been neglected [13] [39, 49]. The users' behaviours may change over time. It follows that their trustworthiness values vary over time; hence, the temporal factor should be assimilated. Moreover, Spammers' behaviours are unsteady as they are not legitimate users although they pretend to be. Therefore, their "temporal patterns of tweeting may vary with frequency, volume, and distribution over time" [50].

Lack of a sentiment analysis of conversations: In the context of social trust, frameworks have been developed to analyse the users' content, taking into consideration the overall feelings regarding what they have chosen to expose their content [17, 19, 51]. However, these efforts did not attempt a sentiment analysis of a post's replies by measuring the trustworthiness values. Sentiments of user's followers are significant to understand the followers' opinions toward the user. Consequently, users who obtain high number positive replies should attain a better reputation than users obtain a large number of negative replies to their content.

***Lack of addressing key features of BD***: BD technology for data storage and analysis provides advanced technical capabilities to the process of analysing massive and extensive data to achieve deep insights in an efficient and scalable manner. Manyika et al. [52] listed some of the Big Data technologies such as Big Table, Cassandra (Open Source DBMS), Cloud Computing, Hadoop (Open Source framework for processing large sets of data), etc. Chen et al., [53] discussed the various open issues and challenges of BD and listed the key technologies of BD. The incorporation of BD technology to facilitate the trustworthiness measuring and inferring tools is unavoidable, especially regarding the nature of the contents of social media which span to a huge quantum. This has interestingly attracted researchers of social trust to leverage the BD techniques to benefit their conducted experiments [54, 55]. However, there has been a lack of addressing the key features of BD such as volume (i.e. massive social data datasets), veracity (i.e. reputation of the data sources), and value (outcome product of the data analytics). Hence, starting from the characteristics of BD and sorting out issues related to these dimensions will be the most efficient way to address BD as well as to benefit trustworthiness analysis endeavors with the expected outcomes of SBD Analysis.

## 4.2 Lack of Managing and Extracting High-Level Domains from the Textual Content of SBD

The prevailing explorations aiming to improve the understanding of the contextual content of social data is noteworthy. Most of the existing approaches to topic distillations rely on bag-of-words techniques such as LDA [33]. However, despite the importance and popularity of these techniques to infer the users' topics of interest, when it comes to context on OSNs there are three main shortcomings; (i) inability to consider the semantic relationships of terms in the user's textual content; (ii) inadequacy to apply its topic modelling technique into short text messages such as tweets; (iii) the high-level topics classifications that use these bag-of-words statistical techniques are inadequate and inferior[28].

LDA extracts latent topics by presenting each topic as a words distribution. This statistical mechanism does not consider the semantic relationships of terms in a document [28]. Furthermore, the high-level topics classifications that use these bag-of-words statistical techniques are inadequate and inferior. Furthermore, using this technique is inappropriate to cluster and search for users based on high-level topics [28]. On the other hand, incorporating semantic web consolidated tools such as AlchemyAPI™ offers a comprehensive list of taxonomies divided into hierarchies where the high-level taxonomy represents the high-level domain and the deeper-level taxonomy provides a fine-grain domain analysis. For instance, "art and entertainment" is considered a high-level taxonomy in which "graphic design" is one of its deep-level taxonomy. LDA is unable to provide high-level topics such as "art and entertainment" from a corpus of posts or tweets unless this term exists in the corpus. Semantic analysis, conversely, extracts semantic concepts and infers high-level domains through analysing the semantic hierarchy of each topic leveraging an ontology, which is not possible using LDA technique.

**4.3   Lack of Domain-based Techniques for Dual Classification.**

This section discusses the mechanisms and limitations of the current approaches to domain-based classification.

***Domain-based classification (inclusion of both user and post levels)***: OSNs have spurred researchers to develop several methods for discovering the main interest(s) of their users. Due to ambiguity, shortness and nosiness of posts such as tweets [28], these endeavours are still immature; Hence, extensive research in this area is vital [56]. For example, Twitter tools [8, 57] are focused on the exploration of user networks to obtain information for user interests and topics. These approaches only extract keywords to obtain a summary of Twitter data. However, the use of keywords only cannot fully cover user domains and may generate misleading user information. There should be incorporation of both user level and tweet level which involve semantics of words and accurate disambiguation for social networks study. The accurate classification of the users' interest assists in providing an accurate understanding of short textual content of future tweets. This benefits several applications the aim of which is to obtain correct domain-based trustworthiness of users and their content in OSNs.

***Incorporation of domain ontology and semantic web, and machine learning***: As indicated earlier, the high-level topics classifications that use the bag-of-words statistical techniques are inadequate, and the brevity and ambiguity of short texts make more difficult the process of topic modelling using these statistical models[58]. Besides, these methods do not consider the temporal factor. In other words, the users' knowledge evolves and their interest might be diverted elsewhere depending on their experience, work, study, etc. Hence, it is important to scrutinise users' interest over time to infer intrinsic topics of interest to users in OSNs. Hence, there is a need for a comprehensive approach which leverages the external domain-based Ontology and semantic web knowledge bases to aid mitigate the disambiguation in the textual content at the user and tweet levels. This approach should be fortified with harnessing advanced and state-of-the-art machine learning techniques to perform domain-based classification at the user and tweet levels.

## 5. Conclusion

Currently, several research efforts are undertaken to handle and manage the large scale of SBD in the quest for added value. A thread of these efforts attempts to provide technical solutions to cope with the volume and speed of social data generation. This is through facilitating the process of data capturing, acquisition and storing. Another host of scholars aim to develop data analytics solutions to enhance the quality of the collected social data, to understand the users' topics of interest, to classify and categorise users into segments, and to extract better insights, thus improving the decision-making process. Despite the significance of these endeavours and initiatives to change the perception of the extracted social data, these current frameworks for SBD analysis only partially consider SBD features. Furthermore, existing efforts are conducted merely to address one or few issues of SBD. A comprehensive framework is required to resolve the issues of data quality, extract hidden knowledge, and infer the credibility of users and their content of the OSNs through extracting the domains of interest at the user and the post level. This consolidates the development of data classification techniques which leads to better anticipating of users' interest in their future published content.

In this paper, an extensive survey of the existing literature is conducted. The discussion of several approaches to measuring the generic-based and domain-based trustworthiness in SBD is carried out followed by addressing the core literature in the areas of semantic analysis, machine learning and data classification within the context of SBD. The carried out efforts within the cordon of these themes are critically reviewed which points out to the considerable achievements have been made in SBD analytics. However, the current approaches and techniques are still insufficient in terms of (i) the lack of domain-based trustworthiness approaches; (ii) the lack of managing and extracting high-level domains from the textual content of SBD; (iii) the lack of domain-based approaches for dual classification of the textual content of SBD.